\documentclass[fleqn,twoside]{article}
\pdfoutput=1
\usepackage[headings]{espcrc2}
\usepackage{amsmath,amssymb}
\usepackage{graphicx}
\usepackage[hyperindex=false]{hyperref}
\usepackage{units}
\usepackage{xspace}
\usepackage[utf8]{inputenc}
\graphicspath{%
  {./}%
}
\newcommand{\Dstar}{\ensuremath{D^{*}}}

\newcommand{\qqbar}{\ensuremath{q\bar{q}}}
\newcommand{\bbbar}{\ensuremath{b\bar{b}}}
\newcommand{\ccbar}{\ensuremath{c\bar{c}}}

\newcommand{\Ftwob}{\ensuremath{F_{2}^{b\bar{b}}}}
\newcommand{\Ftwoc}{\ensuremath{F_{2}^{c\bar{c}}}}

\newcommand{\FLc}{\ensuremath{F_{L}^{c\bar{c}}}}
\newcommand{\Qsq}{\ensuremath{Q^{2}}}
\newcommand{\kT}{\ensuremath{k_{T}}}
\newcommand{\ET}{\ensuremath{E_{T}}}

\newcommand{\pT}{\ensuremath{p_{T}}}
\newcommand{\pTb}{\ensuremath{p_{T}^{b}}}
\newcommand{\pTe}{\ensuremath{p_{T}^{e}}}
\newcommand{\pTrel}{\ensuremath{p_{T}^{\mathrm{rel}}}}
\newcommand{\etae}{\ensuremath{\eta^{e}}}
\newcommand{\dEdx}{\ensuremath{dE/dx}}
\newcommand{\dphi}{\ensuremath{\Delta\phi}}
\newcommand{\signlob}{\ensuremath{\sigma_{b}^{\textrm{NLO}}}}
\newcommand{\signloc}{\ensuremath{\sigma_{c}^{\textrm{NLO}}}}
\newcommand{\sigvisb}{\ensuremath{\sigma_{b}^{\textrm{vis}}}}
\newcommand{\sigvisc}{\ensuremath{\sigma_{c}^{\textrm{vis}}}}
\newcommand{\Ncand}{\ensuremath{N_{\mathrm{cand}}}}
\newcommand{\GeV}{GeV}
\newcommand{\pb}{\mathrm{pb}}
\newcommand{\ipb}{pb$^{-1}$}
\newcommand{\ifb}{fb$^{-1}$}
\newcommand{\PYTHIA}{\textsc{Pythia}}
\newcommand{\CASCADE}{\textsc{Cascade}}
\newcommand{\stat}{\ensuremath{\textrm{stat.}}}
\newcommand{\syst}{\ensuremath{\textrm{syst.}}}
\newcommand{\rnge}{\hbox{$\,\textrm{--}\,$}}

\title{Heavy Quark Production at HERA}
\author{Ian C.\ Brock -- on behalf of the H1 and ZEUS
  Collaborations\\
  Physikalisches Institut, Universität Bonn, Bonn, Germany}
\runtitle{Heavy Quark Production at HERA}
\runauthor{Ian C. Brock}
\begin{document}
\begin{abstract}
  Heavy flavour production is one of the key components of the HERA II
  physics programme. While most of the results presented use leptons
  or the reconstruction of charmed mesons to identify heavy flavour
  production, both the H1 and ZEUS experiments now have working
  microvertex detectors that are being used more and more. In this
  talk I will summarise a selection of the recent results obtained by
  the two collaborations.
\end{abstract}
\maketitle

\section{Introduction}
\label{sec:intro}

The study of heavy flavour production at HERA provides an important
test of perturbative Quantum Chromodynamics (pQCD) and also
valuable information for the measurements to be made at the LHC.
I will discuss a small selection of the many measurements
of heavy flavour production that have been made at HERA,
concentrating on the more recent results.

HERA running started in 1992 and the accelerator stopped running
at the end of June 2007. HERA collided \unit[27.5]{\GeV} electrons or
positrons\footnote{Hereafter unless explicitly stated both electrons
  and positrons are referred to as electrons.} with \unit[920]{\GeV}
(\unit[820]{\GeV} until the end of 1997) protons, giving a
centre-of-mass energy of \unit[318]{\GeV}. A long shutdown in 2000 and
2001 was used to upgrade the machine and the detectors, with the aim
of increasing the luminosity by about a factor of 4. The period up to
2000 is usually called HERA~I and after 2001 HERA~II. By the end of
the running both of the colliding beam experiments, H1 and ZEUS, had
collected about \unit[0.5]{\ifb} of data.

Several kinematic variables are used to characterise the $ep$
scattering process:
\begin{itemize}\setlength{\itemsep}{0pt}
\item \Qsq, the negative squared four-momentum exchanged at the
  electron or positron vertex;
\item $x$, the Bjorken scaling variable;
\item $y$, the inelasticity;
\item $W$, the invariant mass of the hadronic final state.
\end{itemize}

The measurements are usually separated into the deep inelastic
scattering (DIS), $\Qsq \gtrsim \unit[1]{\GeV^{2}}$, or
photoproduction, $\Qsq \lesssim \unit[1]{\GeV^{2}}$, regimes, depending
on whether the scattered electron is detected in the in the main
calorimeter or not.

\begin{figure}[htbp]
  \centering
  \includegraphics[width=0.6\columnwidth]{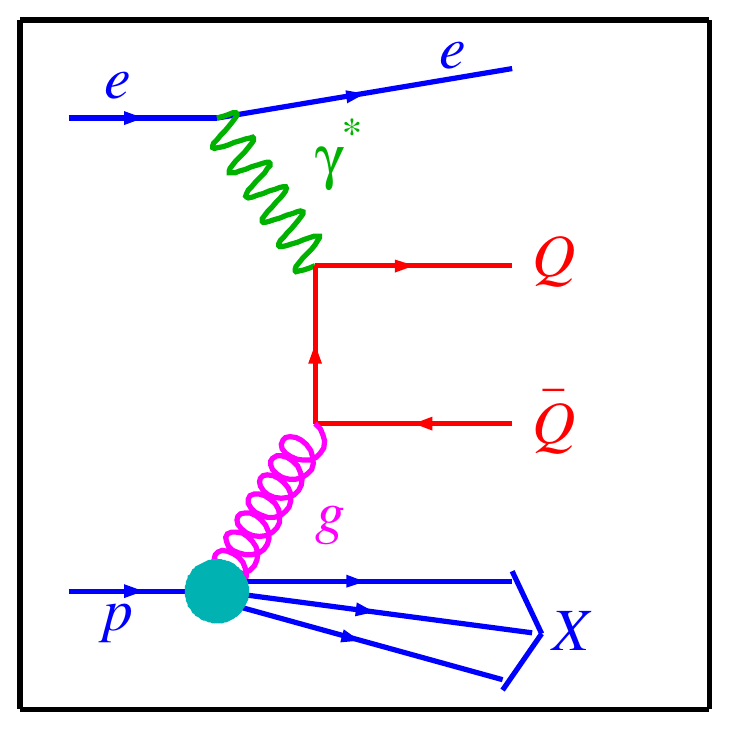}
  \caption{Feynman diagram of the boson gluon fusion process.}
  \label{fig:bgf}
\end{figure}

The main production mechanism for heavy quarks is the
so-called Boson Gluon Fusion (BGF) process which is illustrated in
Figure~\ref{fig:bgf}. 
In practice higher order contributions also have to be taken into
account. The photon can fluctuate into a \qqbar{} pair and one of the
quarks then participates in the hard interaction (resolved
photoproduction); or in so-called excitation the scattering from charm
or beauty can take place with intrinsic charm or beauty inside the
proton or photon. Monte Carlo models usually include these process in
addition to the direct boson-gluon fusion. These higher order
processes are also sometimes referred to collectively as non-direct.

Next-to-leading oder (NLO) QCD calculations exist for heavy quark
production at HERA. These are implemented for photoproduction in the
FMNR programme \cite{np:b412:225,asdhep:15:609} and for DIS in the
HVQDIS programme \cite{pr:d57:2806}. The programmes include simple
independent fragmentation of the $b$ or $c$ quark, but on their own
are not able to give predictions for correlations between final-state
particles. The FMNR programme has been interfaced \cite{Geiser:2007py}
to the \PYTHIA{} Monte Carlo and its predictions have been used for
studies of dimuon final states.

Most heavy flavour measurements rely on the central tracking detectors
and are helped significantly by the presence of a microvertex
detector. H1 had such a detector for some of the HERA~I running, while
both ZEUS and H1 had such detectors for the HERA~II running period.

Several different methods have been used by the collaborations to
identify the production of heavy quarks. Each of them has its
advantages and disadvantages and often cover different kinematic
ranges. Traditionally the identification of \Dstar{} mesons and
semileptonic decays have been used most often. With the advent of
microvertex detectors lifetime information is being used more and more
often. This works best for events with high energy jets, while tagging
with semileptonic decays to electrons or double tags enables one to go
to lower transverse momenta.

\section{\Dstar{} Production}
\label{sec:dstar}

\begin{figure}[htbp]
  \centering
  \includegraphics[width=0.9\columnwidth]{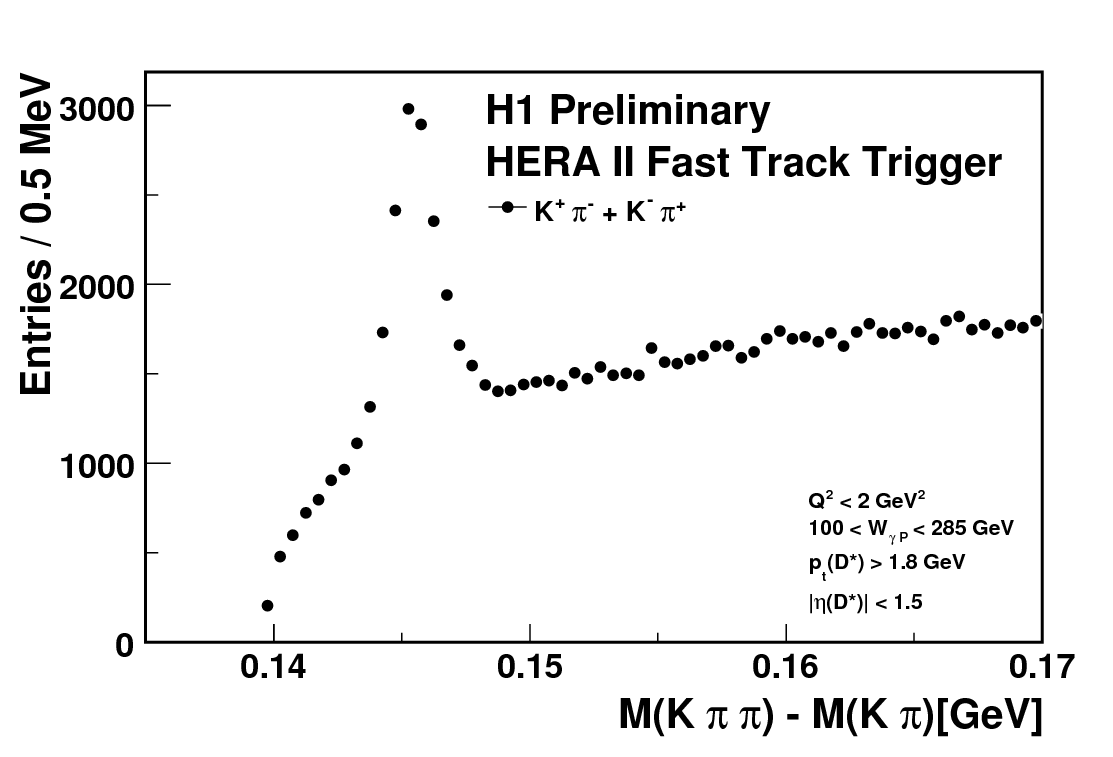}\\
  \includegraphics[width=0.9\columnwidth]{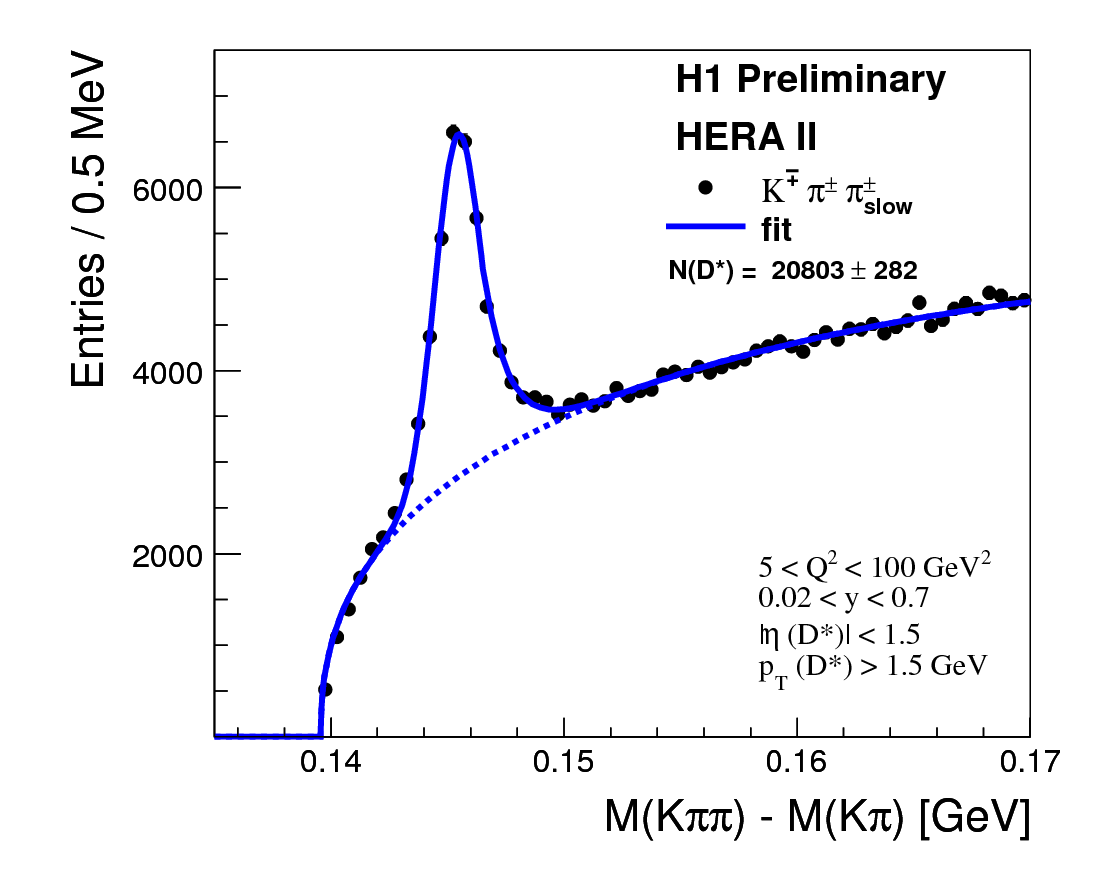}
  \caption{$\Delta M = m(K\pi\pi) - m(K\pi)$ distributions for the
    photoproduction (top) and DIS (bottom) samples. The points show
    the data, the curves in the lower figure, the results of a fit to
    the distribution. The dotted line shows the background shape. The
    kinematic cuts are indicated.}
  \label{fig:dstar-peak}
\end{figure}

%
%
The H1 collaboration has made a series of measurements of \Dstar{}
production both in the photoproduction \cite{H1:DIS08:DstarPhP}
(\unit[96]{\ipb} of data taken in 2006 and 2007) and the
DIS \cite{h1:DIS08:DstarDIS} (\unit[347]{\ipb} from the HERA~II running
period) regimes. For these measurements they make use of their new
Fast Track Trigger, which enables events with \Dstar{} candidates to
be selected early in the trigger chain. In both photoproduction and
DIS clear signals are seen as illustrated in
Figure~\ref{fig:dstar-peak}.  The kinematic cuts are indicated in the
figures.

\begin{figure}[htbp]
  \centering
  \includegraphics[width=0.8\columnwidth]{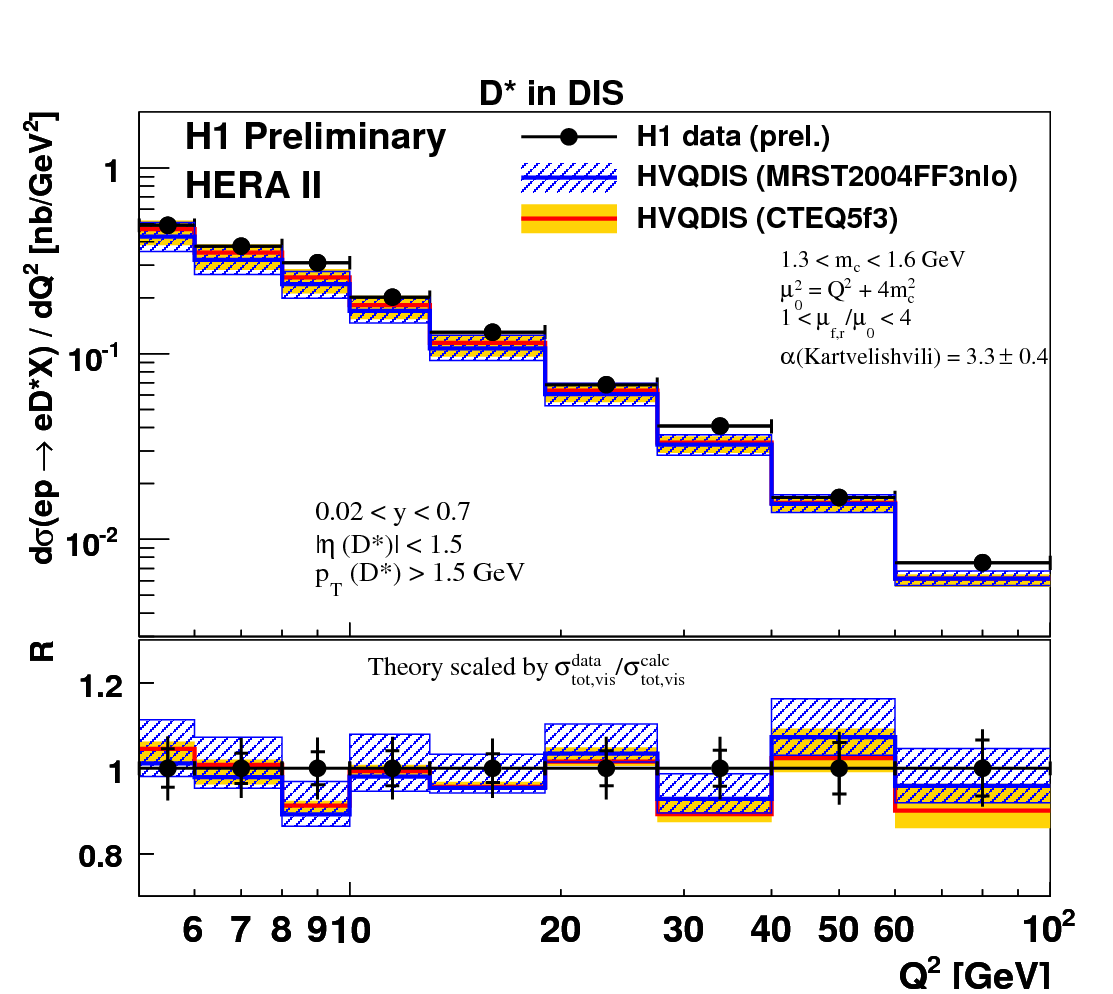}
  \caption{\Dstar{} production cross-section as a function of
    \Qsq. The data (points) are compared to NLO QCD predictions with
    two different PDFs. The lower plot shows the NLO QCD prediction
    scaled by the ratio of the data to NLO QCD visible
    cross-sections. The range of variation of the quark mass and
    factorisation and renormalisation scales are indicated in the
    figure.}
  \label{fig:dstar-qsq}
\end{figure}

Differential cross-sections as a function of a wide range of variables
have been determined. The cross-section as a function of \Qsq{} is
shown in Figure~\ref{fig:dstar-qsq}.
The measurements are compared to the predictions of the HVQDIS
programme using two different Parton Distribution Functions (PDF). The
different PDFs show a very similar behaviour as a function of \Qsq.

\begin{figure}[htbp]
  \centering
  \includegraphics[width=0.85\columnwidth]{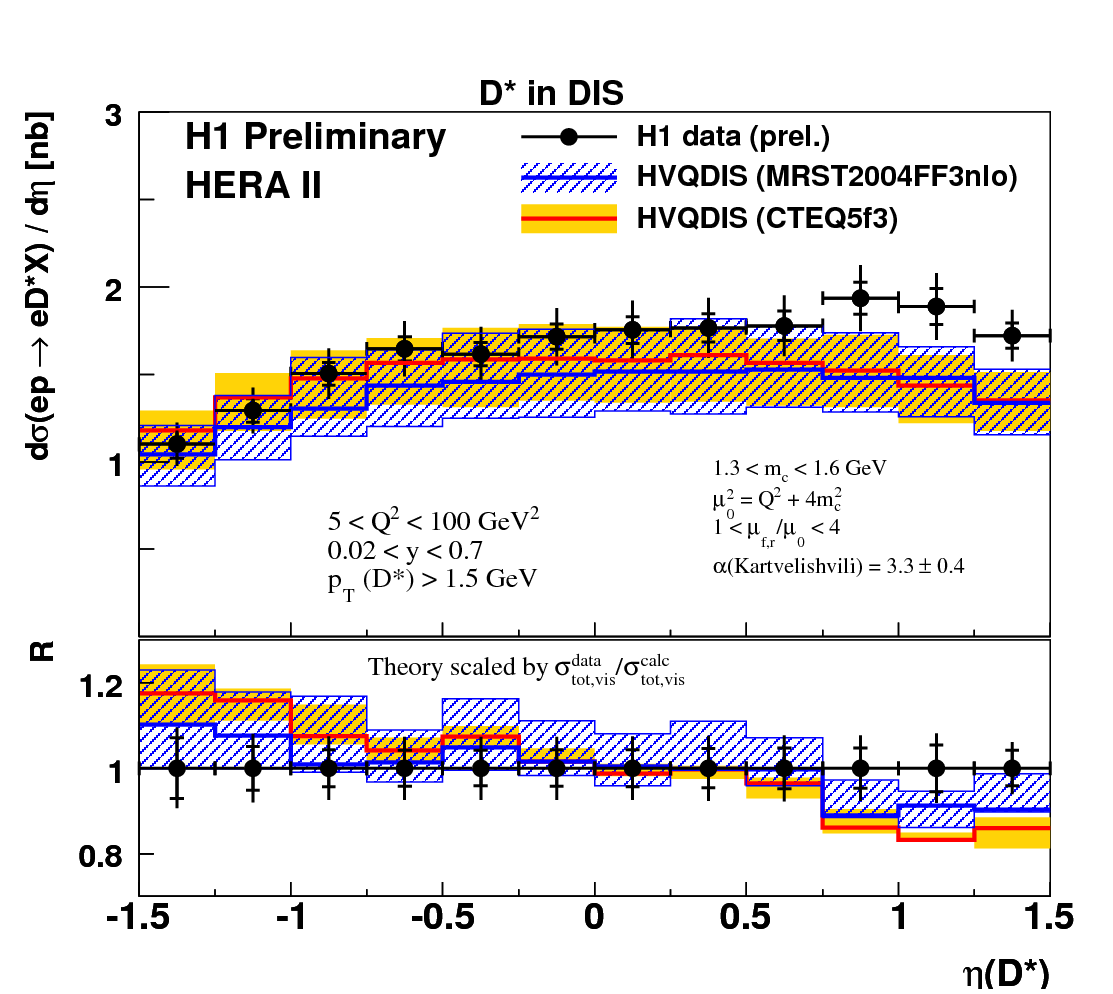}
  \caption{\Dstar{} production cross-section in DIS as a function of
    $\eta(\Dstar)$ compared to NLO QCD predictions. For further
    details see the caption of Figure~\protect\ref{fig:dstar-qsq}.
  }
  \label{fig:dstar-eta-dis}
\end{figure}

In contrast the cross-section as a function of the pseudorapidity,
$\eta$, shows a clear dependence (see Figure~\ref{fig:dstar-eta-dis}),
demonstrating the sensitivity of charm production to the gluon
structure function.

\begin{figure}[htbp]
  \centering
  \includegraphics[width=0.85\columnwidth]{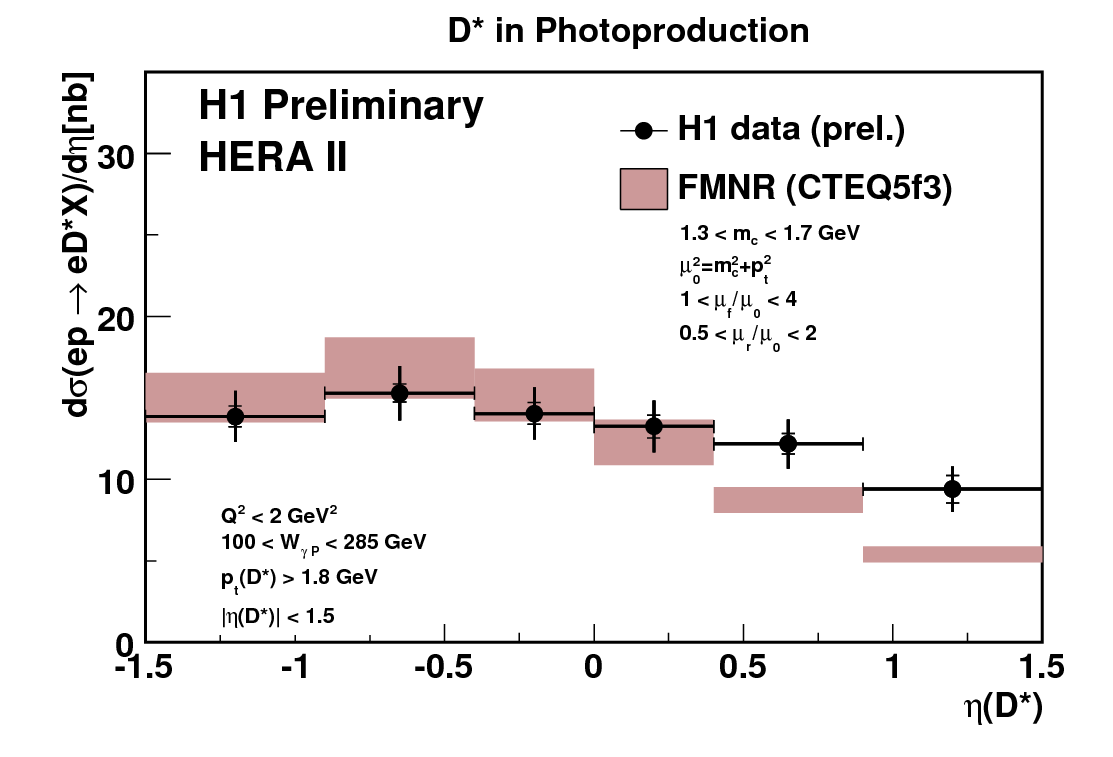}
  \caption{\Dstar{} production cross-section in photoproduction as a
    function of $\eta(\Dstar)$ compared to the NLO QCD prediction. The
    band shows the uncertainty in the prediction. The range of variation
    of the quark mass and factorisation and renormalisation scales
    are indicated in the figure.}
  \label{fig:dstar-eta-php}
\end{figure}

\begin{figure}[htbp]
  \centering
  \includegraphics[width=0.85\columnwidth]{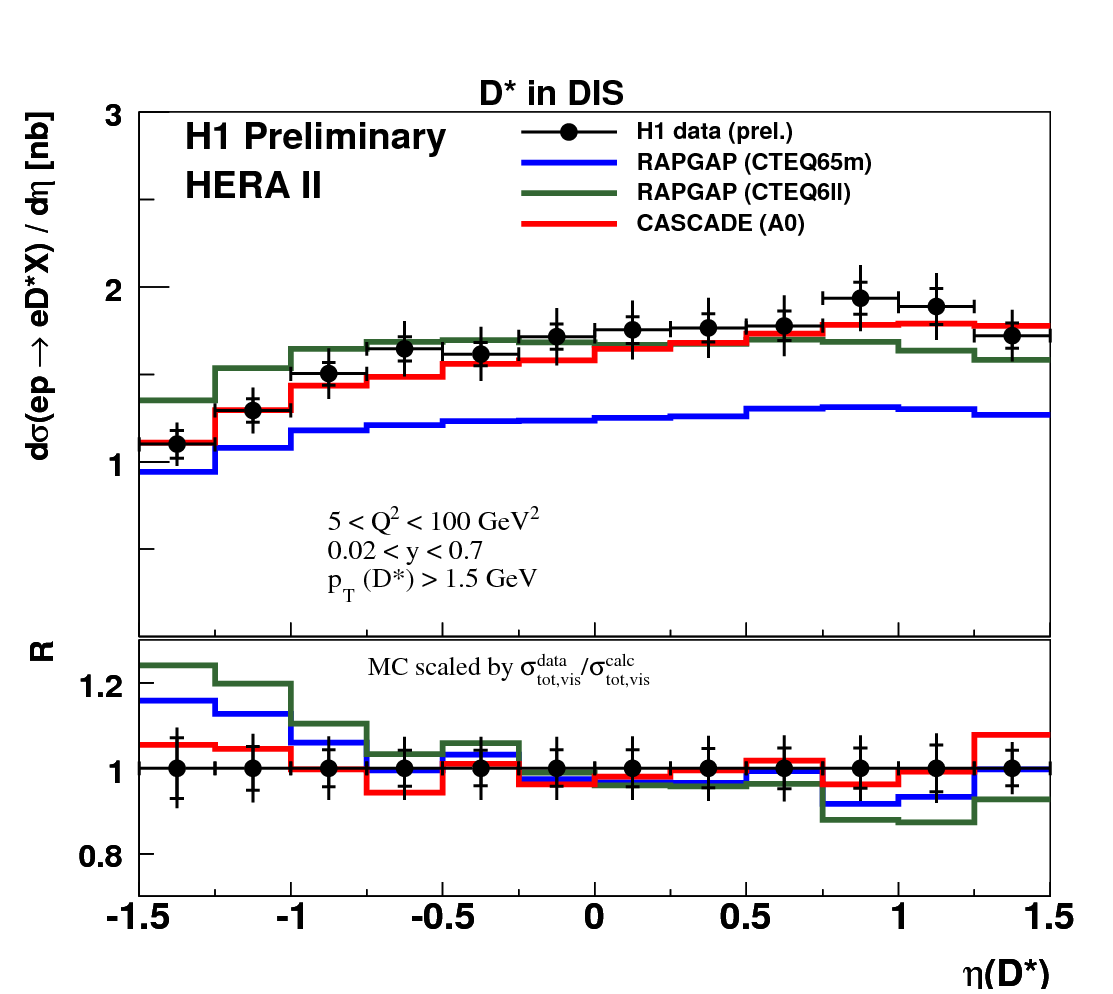}
  \caption{\Dstar{} production cross-section in DIS as a
    function of $\eta(\Dstar)$. The data are compared  to the
    different Monte Carlo generators as indicated in the figure.}
  \label{fig:dstar-eta-phpmc}
\end{figure}

Making the same comparison for photoproduction one sees that the
NLO QCD prediction underestimates the data in the forward direction
(see Figure~\ref{fig:dstar-eta-php}).
Comparing the data to Monte Carlo predictions, the \CASCADE{}
generator, which is based upon \kT-factorisation, is able to provide a
better description of the data in DIS than RAPGAP with two different
PDFs (see Figure~\ref{fig:dstar-eta-phpmc}).
However, in photoproduction \PYTHIA, using the massless scheme for
the generation of the heavy quarks, provides the best description of
the data, while \CASCADE{} and \PYTHIA{} in the massive mode both
undershoot the data in the forward direction.

\section{Beauty in Photoproduction}
\label{sec:bsl}

%
%
The ZEUS collaboration recently reported two new measurements of
$b$-quark production in photoproduction.
The first measurement uses semi\-leptonic decays to
muons \cite{ZEUS:EPS07:Btomu} to identify heavy quark decays, while the
second one uses electrons \cite{Chekanov:2008aa}. Dijet events are selected
by requiring at least two jets with $|\eta| < 2.5$ and a transverse
momentum (muon measurement) or energy (electron measurement) greater
than \unit[7]{\GeV} for the highest transverse energy and
\unit[6]{\GeV} for the 2nd highest transverse energy jet.

\begin{figure}[htbp]
  \centering
  \includegraphics[width=0.80\columnwidth]{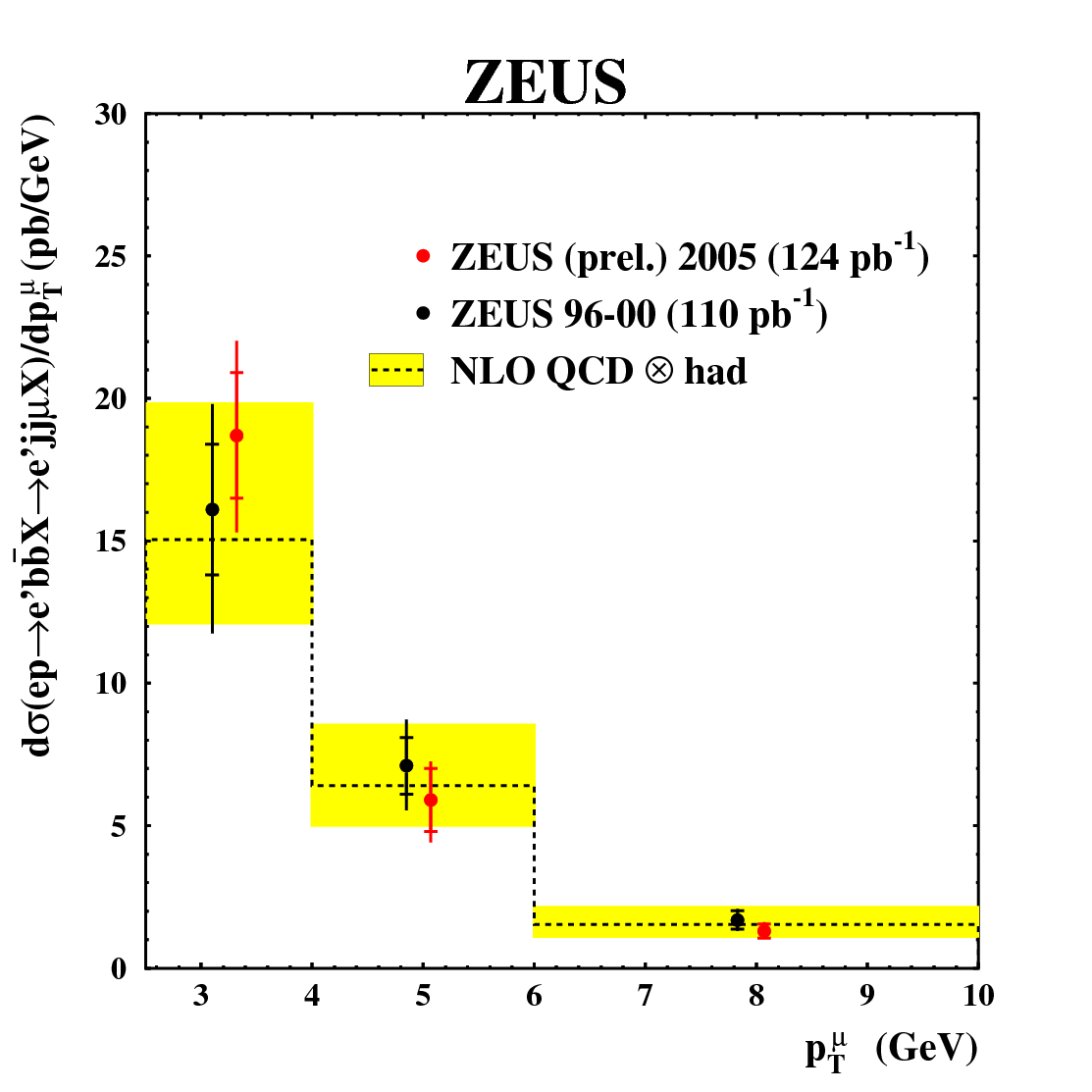}
  \caption{$b$-quark production cross-section in dijet photoproduction as a
    function of the transverse momentum of the
    muon. The measurement is compared to an earlier ZEUS
    publication \protect\cite{pr:d70:012008} and the NLO QCD
    prediction. The uncertainty on the prediction is indicated by the
    band.}
  \label{fig:btomu-xsect-ptmu}
\end{figure}

\begin{figure}[htbp]
  \centering
  \includegraphics[width=0.80\columnwidth]{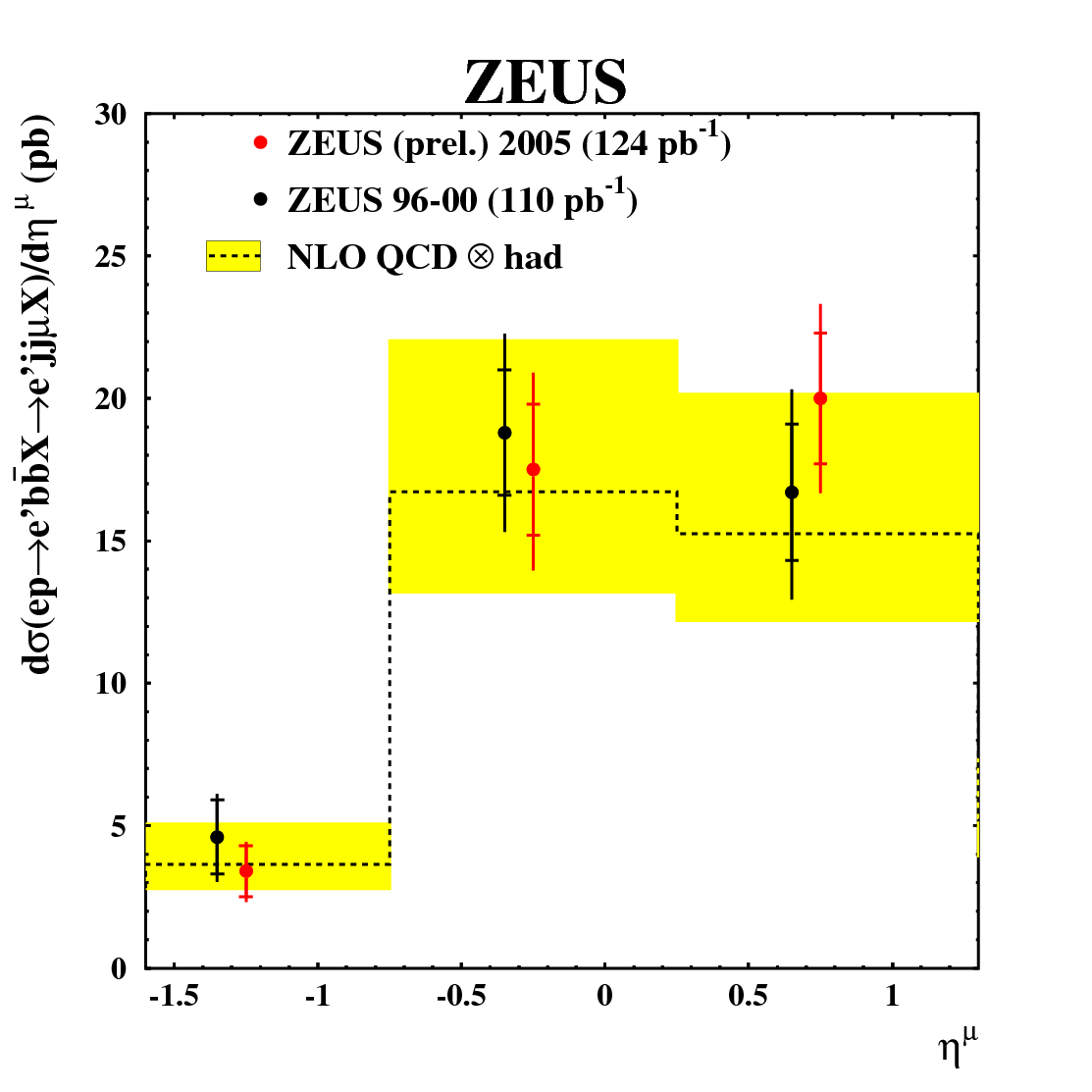}
  \caption{$b$-quark production cross-section in dijet photoproduction as a
    function of the pseudorapidity of the muon. The measurement is
    compared to an earlier ZEUS
    publication \protect\cite{pr:d70:012008} and the NLO QCD
    prediction. The uncertainty on the prediction is indicated by the
    band.}
  \label{fig:btomu-xsect-etamu}
\end{figure}

The first measurement uses \unit[124]{\ipb} of data collected in 2005
and requires a well identified muon with $\pT >
\unit[2.5]{\GeV}$ and $-1.6 < \eta < 2.3$.  The
microvertex detector is used to measure the impact parameter
($\delta$) of identified muons. This is combined with the transverse
momentum of the muon with respect to the jet axis (\pTrel) to separate
$b$-quark events from background. The cross-sections as a function of
the transverse momentum and pseudorapidity of the muon are shown in
Figures~\ref{fig:btomu-xsect-ptmu} and \ref{fig:btomu-xsect-etamu}.
Both the charm and beauty contributions are left free in the fit.
The figures also show the results of an analysis using the
HERA~I data \cite{pr:d70:012008}, which only used \pTrel{} for
separation. For this earlier analysis the charm content was
fixed. Good agreement between the analyses is seen. The predictions of
the NLO QCD prediction are also in good agreement with the data.

\begin{figure}[htbp]
  \centering
  \includegraphics[width=0.60\columnwidth]{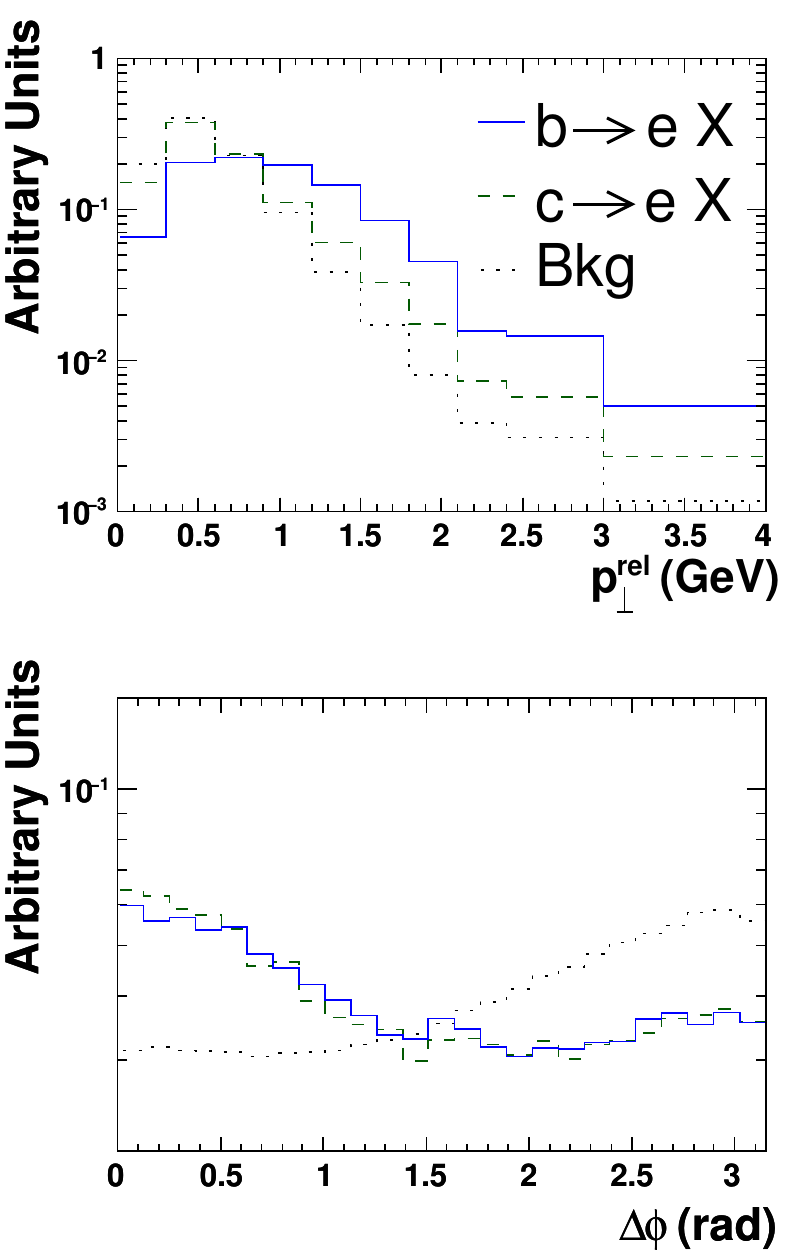}
  \caption{Variables used to separate semileptonic heavy quark decays
    from background.  The solid line shows the distribution for
    electrons from semileptonic $b$-quark decays, the dashed line for
    $c$-quark decays and the dotted line the background (Bkg).}
  \label{fig:btoe-vars}
\end{figure}

The second measurement identifies electrons from the semi\-leptonic
decays of heavy quarks using an integrated luminosity of
\unit[120]{\ipb} collected during the HERA~I running period.  Electron
identification uses the measurement of the specific energy loss,
\dEdx, in the Central Tracking Detector, CTD, the fraction of the
energy deposited in the electromagnetic calorimeter as well as the
ratio of the energy deposited in the calorimeter to the momentum
measured in the tracking detectors. Semileptonic decays are separated
from background using \pTrel{} and the azimuthal angle between the
electron direction and the missing transverse momentum vector,
$\dphi$. As illustrated in Figure~\ref{fig:btoe-vars}, \pTrel{} can
separate $b$-quark from $c$-quark decays, while \dphi{} separates
semileptonic decays from background.

\begin{figure}[htbp]
  \centering
  \includegraphics[width=0.70\columnwidth]{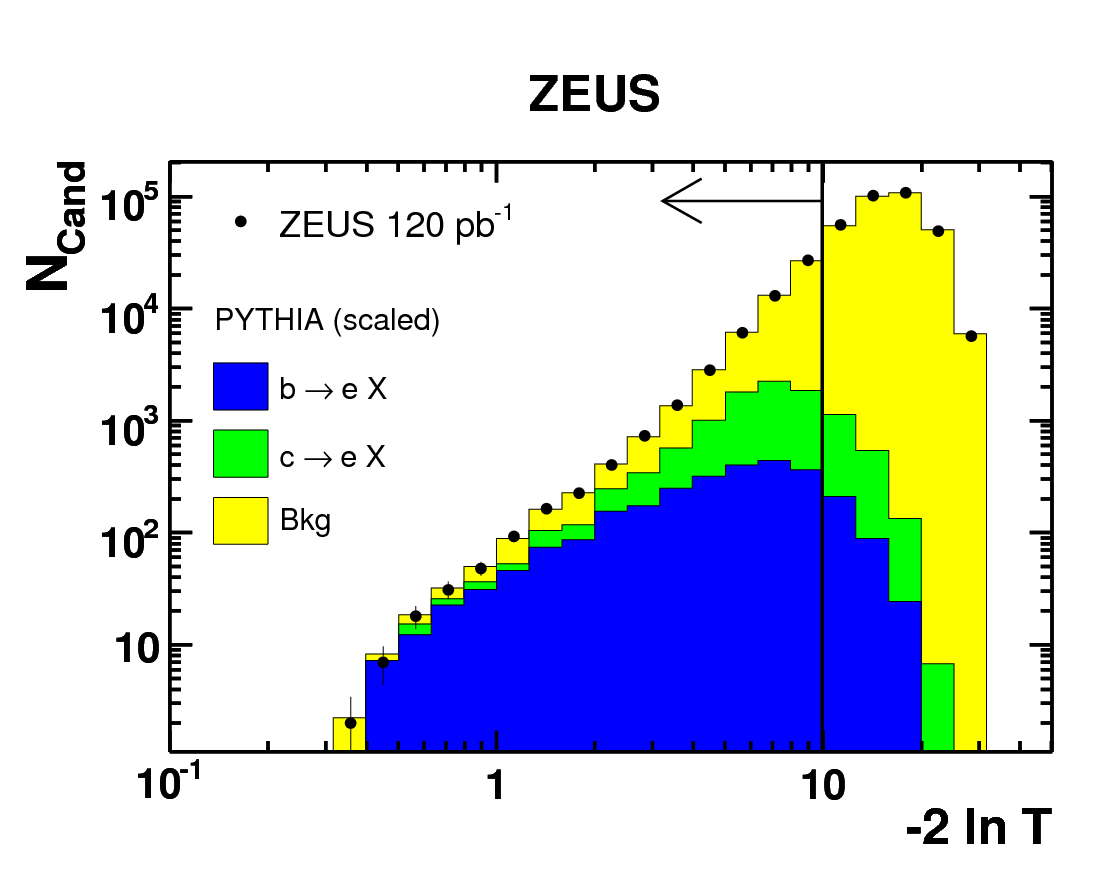}
  \caption{The distribution of the likelihood ratio for electron
    candidates, \Ncand, in data compared to the Monte Carlo
    expectation after the scaling the predictions to best fit the
    data.  The arrow indicates the region included in the fit ($-2 \ln
    T < 10)$.  The shaded areas show the fitted contributions from $b$
    quarks, $c$ quarks and background as denoted in the figure.}
  \label{fig:btoe-like}
\end{figure}

The variables are combined using a likelihood ratio method, which is
optimised for the identification of electrons from semileptonic
$b$-quark decays. The distribution of the likelihood ratio is shown in
Figure~\ref{fig:btoe-like}. The distribution is fit using the expected
distributions for beauty, charm and background to determine the
fractions of events from each source. The fit provides a very good
description of the data.

The visible $ep$ cross sections (at hadron level) for $b$-quark
and $c$-quark production and the subsequent semileptonic decay to an
electron with $\pTe > \unit[0.9]{\GeV}$ in the range $|\etae| < 1.5$
in photoproduction events with $\Qsq < \unit[1]{\GeV^{2}}$ and $0.2 <
y < 0.8$ and at least two jets with $\ET > \unit[7 (6)]{\GeV}$,
$|\eta| < 2.5$ were determined separately for
$\sqrt{s}=\unit[300]{\GeV}$ and $\sqrt{s}=\unit[318]{\GeV}$.  For the
complete data set ($96 \rnge 00$) the cross-sections evaluated at
$\sqrt{s}=\unit[318]{\GeV}$ are
\begin{align*}
\sigvisb & = 
\left( 125 \pm 11 (\stat) ^{+10}_{-11} (\syst) \right)\pb,\\ 
\sigvisc & = 
\left( 278 \pm 33 (\stat) ^{+48}_{-24} (\syst) \right)\pb.
\end{align*}
These cross-sections are in agreement with the corresponding NLO QCD
predictions:
\begin{align*}
\signlob & = 
\left( 88 ^{+22}_{-13} \right)\pb,\\ 
\signloc & = 
\left( 380 ^{+48}_{-24} \right)\pb.
\end{align*}

\begin{figure}[htbp]
  \centering
  \includegraphics[width=0.9\columnwidth]{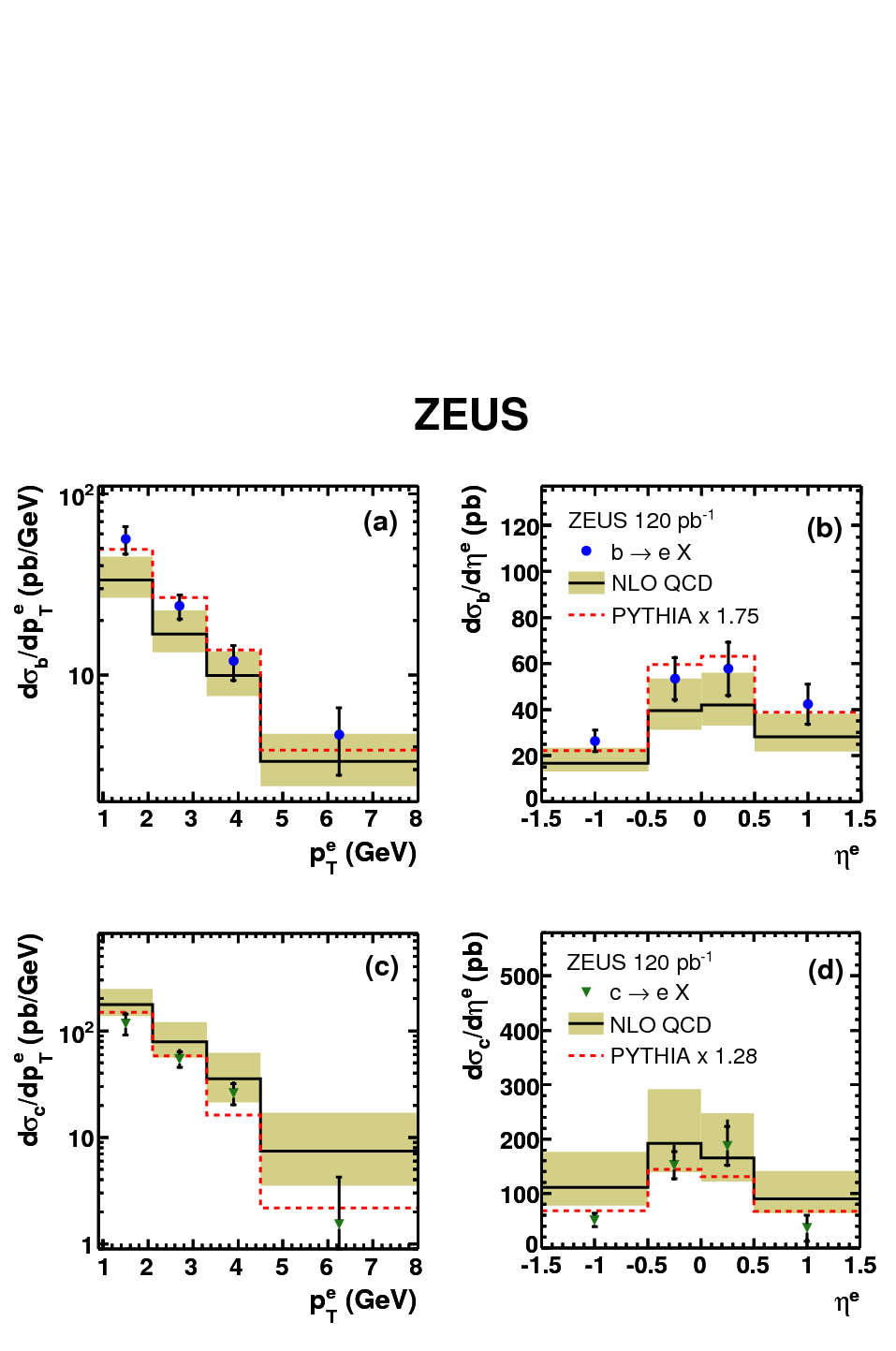}
  \caption{Differential cross sections  as a function of 
    a), c) the transverse momentum
    and b), d) the pseudorapidity of the electrons. Plots a) and b) are for
    $b$-quark production while c) and d) are for $c$-quark
    production.
    The measurements are shown as points. The inner error bar shows the
    statistical uncertainty and the outer error bar shows the statistical and
    systematic uncertainties added in quadrature.
    The solid line shows the NLO QCD prediction after hadronisation corrections, 
    with the theoretical uncertainties indicated by the band; the
    dashed line shows the scaled prediction from \PYTHIA.
  }
  \label{fig:btoe-xsect}
\end{figure}

The cross-sections as a function of the transverse momentum and
pseudorapidity of the electron are shown in
Figure~\ref{fig:btoe-xsect}.
Good agreement with the NLO QCD prediction is seen. The data also agree
well with the \PYTHIA{} prediction scaled by a factor of 1.75 for
$b$-quark production and 1.28 for $c$-quark production.

\section{Beauty Correlations}
\label{sec:bcorr}

The identification of both heavy-quark decays in an event has several
advantages: the background is reduced substantially and the kinematic
range accessible is larger. The disadvantage is a significant
reduction of statistics. If both $b$-quark jets can be identified,
dijet correlations can be directly measured which probe
next-to-leading order effects. The ZEUS collaboration used the HERA~I
data sample to select events with $\ET > \unit[8]{\GeV}$ and two
identified muons \cite{thesis:bloch:2005}.  Separating the samples
according to the relative charge of the muons as well as their
invariant mass allows much of the background to be evaluated directly
from the data. 

\begin{figure}[htbp]
  \centering
  \includegraphics[width=0.90\columnwidth]{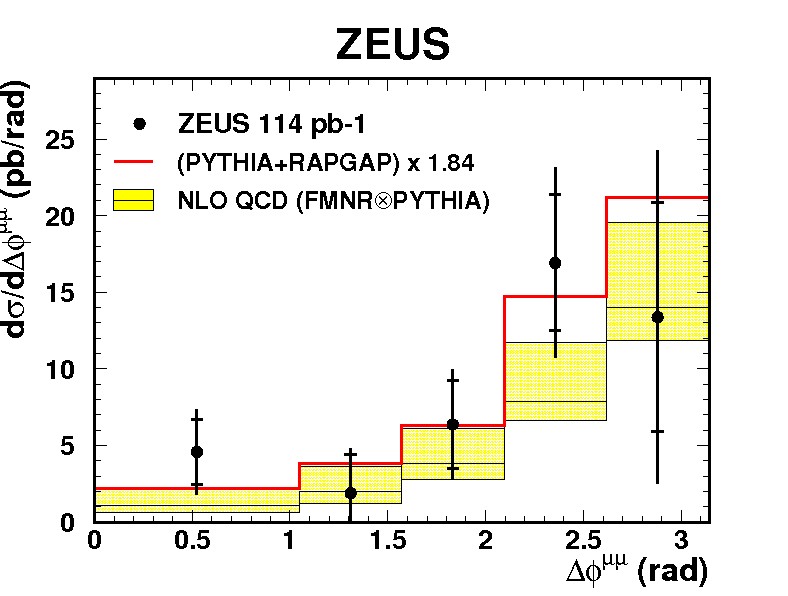}
  \caption{The dimuon cross-section as a function of the azimuthal
    angle between the muons in dijet events. The data (points) are
    compared to the NLO QCD prediction calculated using the
    FMNRxPYTHIA interface as well as to the RAPGAP prediction scaled
    by a factor of 1.84. The inner error bar shows the statistical
    uncertainty and the outer error bar shows the statistical and
    systematic uncertainties added in quadrature.
  }
  \label{fig:bcorr-dphi2}
\end{figure}

The extracted cross-section is shown in Figure~\ref{fig:bcorr-dphi2}.
It is compared with the NLO QCD prediction as well as with a leading
order Monte Carlo. Given the limited statistics it is not possible to
say whether the NLO calculation provides a better description of the
data. 

\section{\Ftwob{} and \Ftwoc}
\label{sec:f2bc}

\begin{figure}[htbp]
  \centering
  \includegraphics[width=0.80\columnwidth]{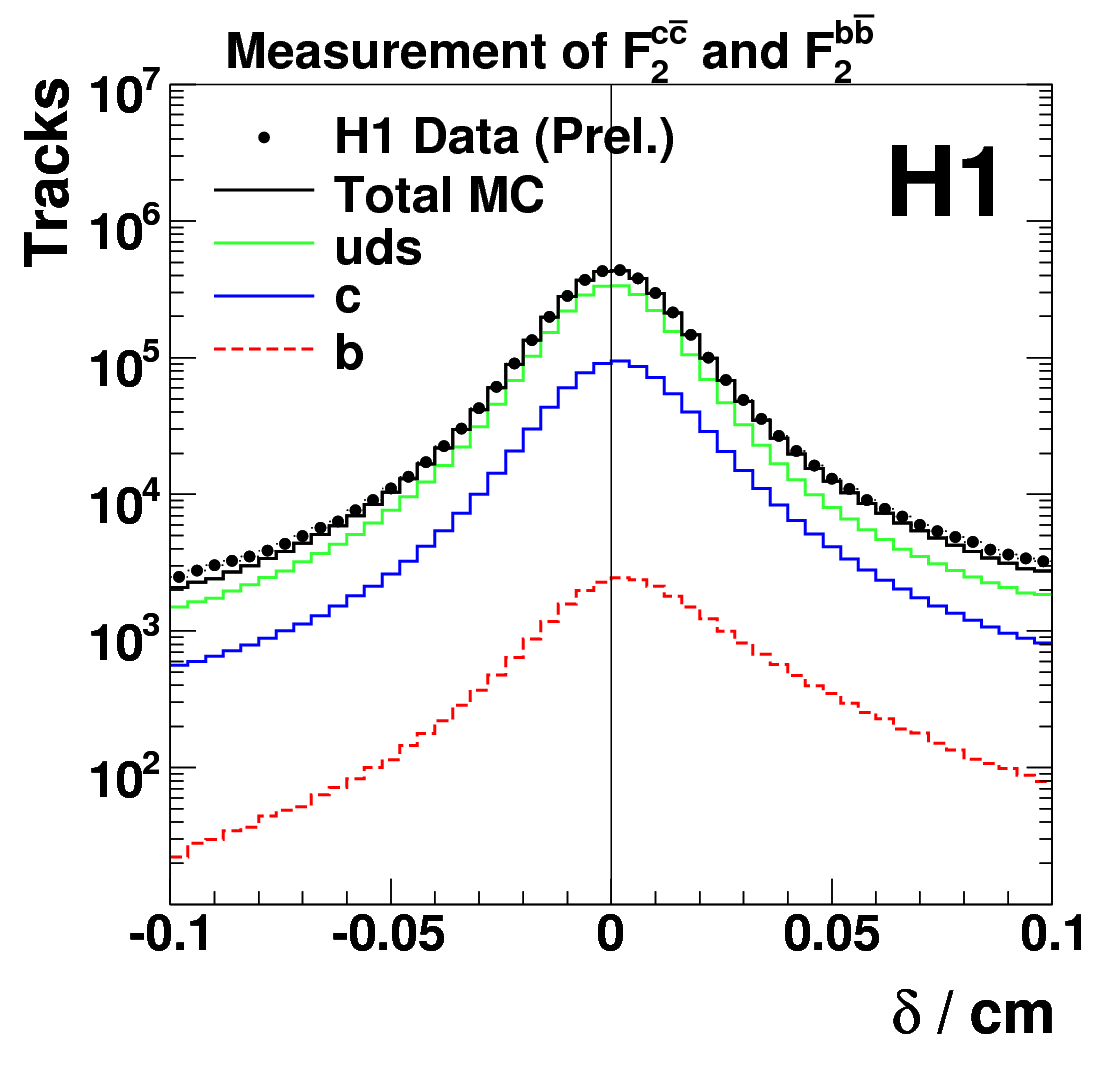}
  \caption{Distribution of the signed impact parameter for the data
    (points) and the contributions from $b$, $c$ and light
    quarks, evaluated using Monte Carlo events.}
  \label{fig:f2bc-ip}
\end{figure}

The H1 collaboration have used the impact parameter significance to
determine \Ftwob{} and \Ftwoc{} for $ \unit[12]{\GeV^{2}} < \Qsq <
\unit[650]{\GeV^{2}}$ and $0.0002 < x < 0.032$ using \unit[56]{\ipb}
of data taken in 2006 \cite{H1:LP07:F2bc}. The distribution of the
impact parameter in the transverse plane is shown in
Figure~\ref{fig:f2bc-ip}.

In order to separate beauty and charm from background the two tracks
with the most significant impact parameters ($S_{1}$ and $S_{2}$
respectively) are selected, rejecting events where the signs of the
impact parameters differ. The significance is defined as $\delta /
\sigma(\delta)$, where $\sigma(\delta)$ is the error on
$\delta$. Events with one good track are used to make the $S_{1}$
distribution; all other events are used for the $S_{2}$ distribution.

\begin{figure}[htbp]
  \centering
  \includegraphics[width=0.80\columnwidth]{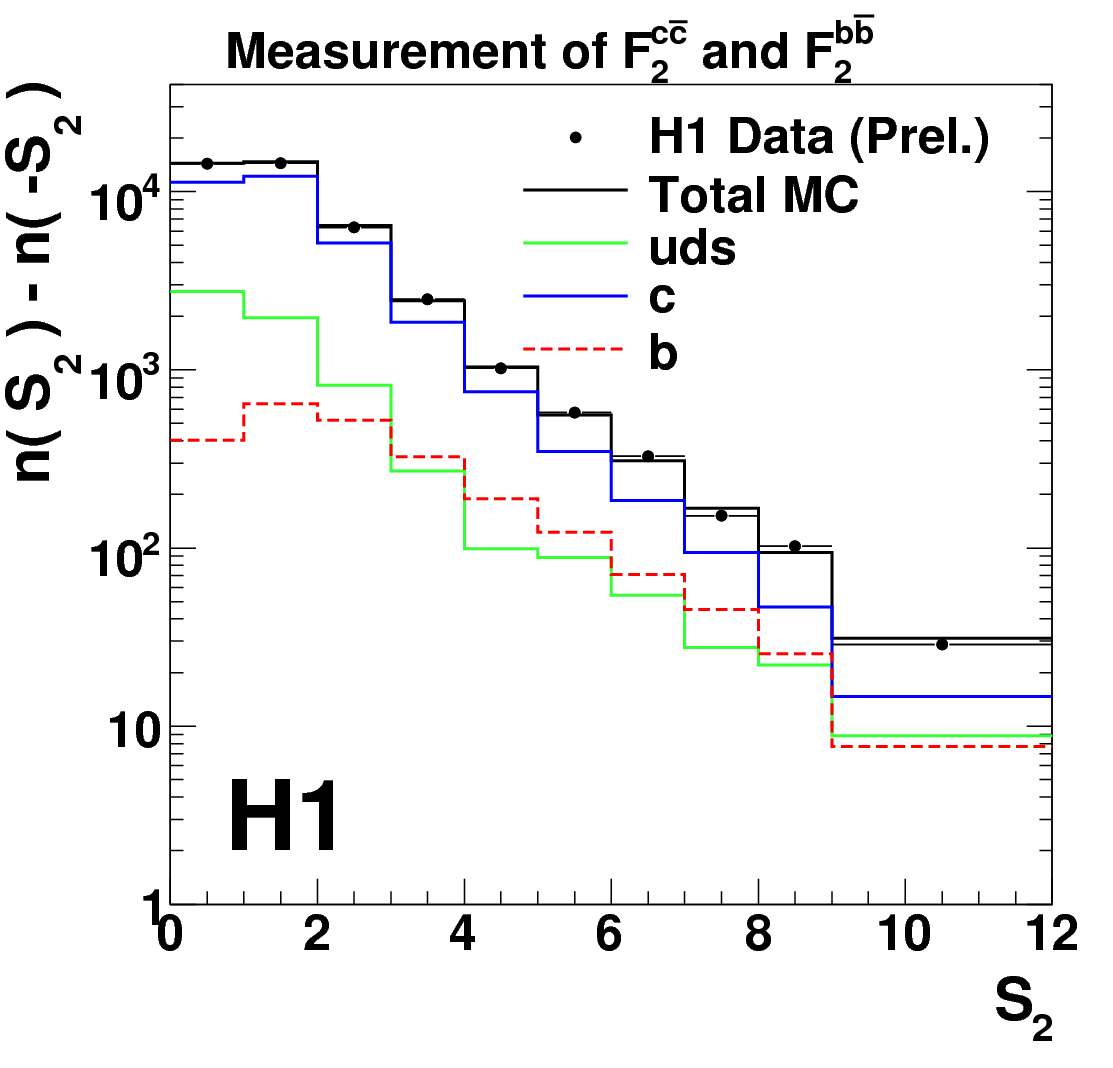}
  \caption{Distribution of the second highest significance and the
    contributions from $b$, $c$ and light quarks, evaluated using
    Monte Carlo events.}
  \label{fig:f2bc-sig}
\end{figure}

\begin{figure}[tbp]
  \centering
  \includegraphics[width=0.9\columnwidth]{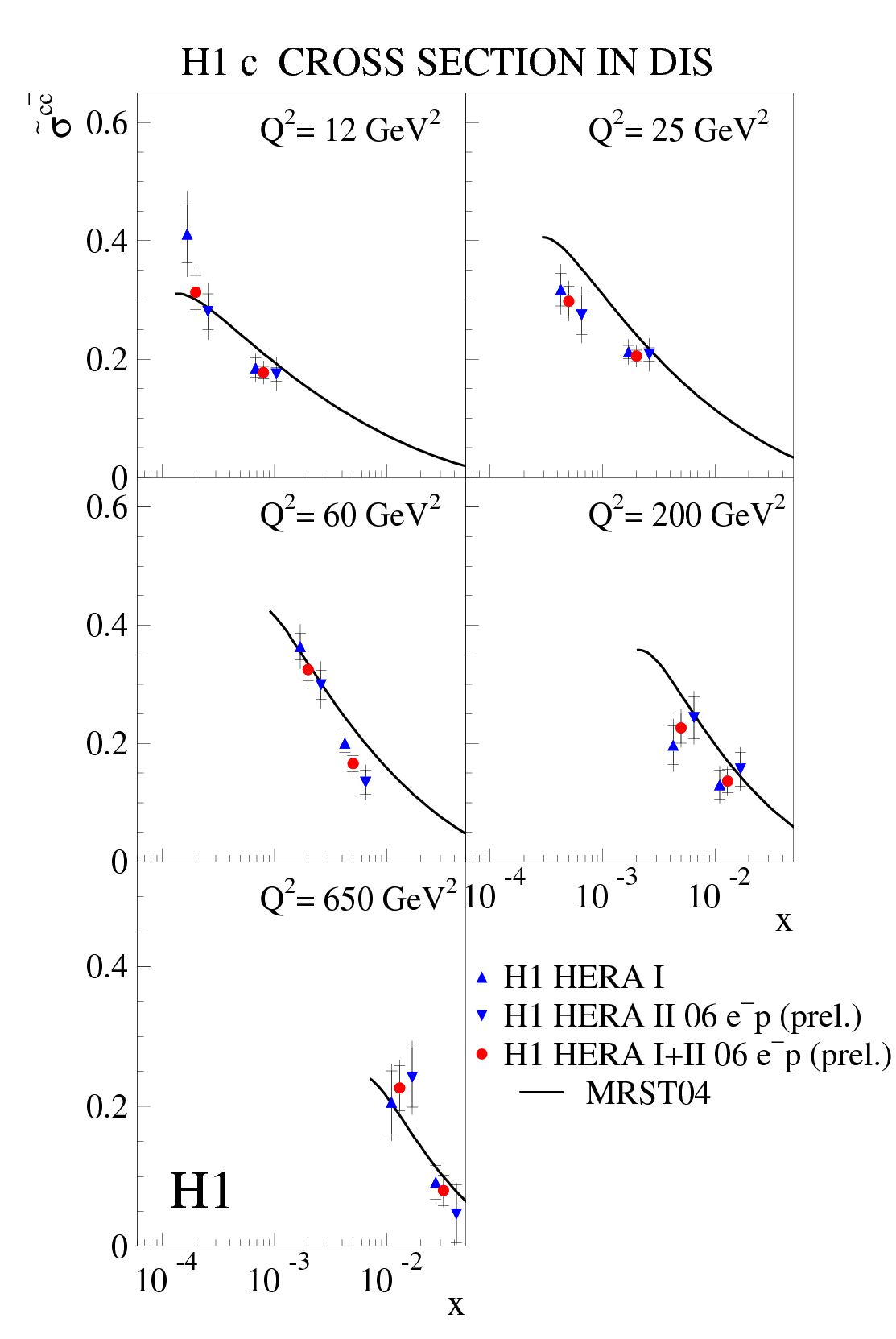}
  \caption{The reduced cross-section for \ccbar{} production in
    different \Qsq{} bins as a function of $x$. The inner errors bars
    show the statistical uncertainty, the outer error bars represent
    the statistical and systematic uncertainties added in
    quadrature. The measurements are compared with those from HERA~I
    and the averaged data are also shown. The measurements are
    compared to the MRST04 prediction.}
  \label{fig:f2bc-xsectr}
\end{figure}

The contents of the negative significance bins are subtracted from the
corresponding positive significance bins.  This yields the
distribution of $S_{2}$ shown in Figure~\ref{fig:f2bc-sig}.
The charm contribution dominates the distribution. At high
significance the beauty contribution becomes larger. The data are
split into bins in $\Qsq$ and $x$ and the contributions of beauty and
charm are determined separately in each bin using a least squares
simultaneous fit to the $S_{1}$ and $S_{2}$ distributions. The overall
normalisation is determined by also including in the fit the total
number of inclusive events without any cut on the impact parameter
significance. The results of the fit in each $x - \Qsq$ bin are
converted to a ``reduced cross-section'' using
\begin{equation*}
  \tilde{\sigma}^{\ccbar}(x, \Qsq) =
  \frac{d^{2}\sigma^{\ccbar}}{dx\,d\Qsq}
  \frac{x Q^{4}}{2 \pi \alpha^{2} (1 + (1-y)^{2})}\,.
\end{equation*}
The reduced cross section for \ccbar{}
production as a function of $x$ in different \Qsq{} bins is shown in
Figure~\ref{fig:f2bc-xsectr}.
Measurements made with the HERA~I data in the same kinematic range are
also shown in the figure. The results are in good agreement with each
other and can be combined, taking into account the correlated
systematic errors. The prediction using the Variable Flavour Number
Scheme (VFNS) in the MRST04 PDF agrees well with the data.

\begin{figure}[htbp]
  \centering
  \includegraphics[width=\columnwidth]{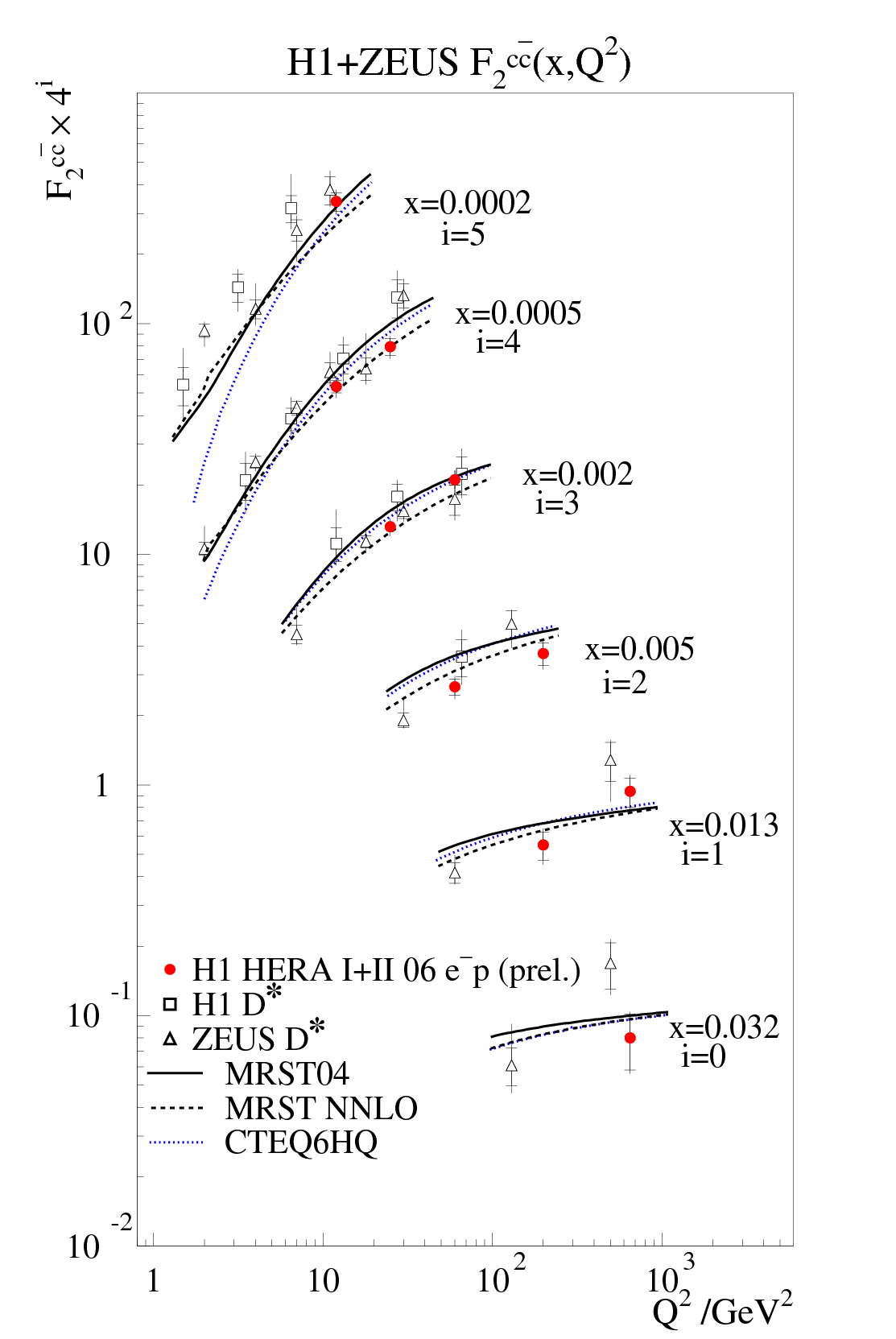}
  \caption{\Ftwoc{} as a function of \Qsq{} for different $x$
    ranges. Also shown are measurements from the H1 and ZEUS collaborations
    using \Dstar{} mesons to identify the charm quarks. The inner errors bars
    show the statistical uncertainty, the outer error bars represent
    the statistical and systematic uncertainties added in
    quadrature.
  }
  \label{fig:f2bc-f2c}
\end{figure}

\begin{figure}[htbp]
  \centering
  \includegraphics[width=\columnwidth]{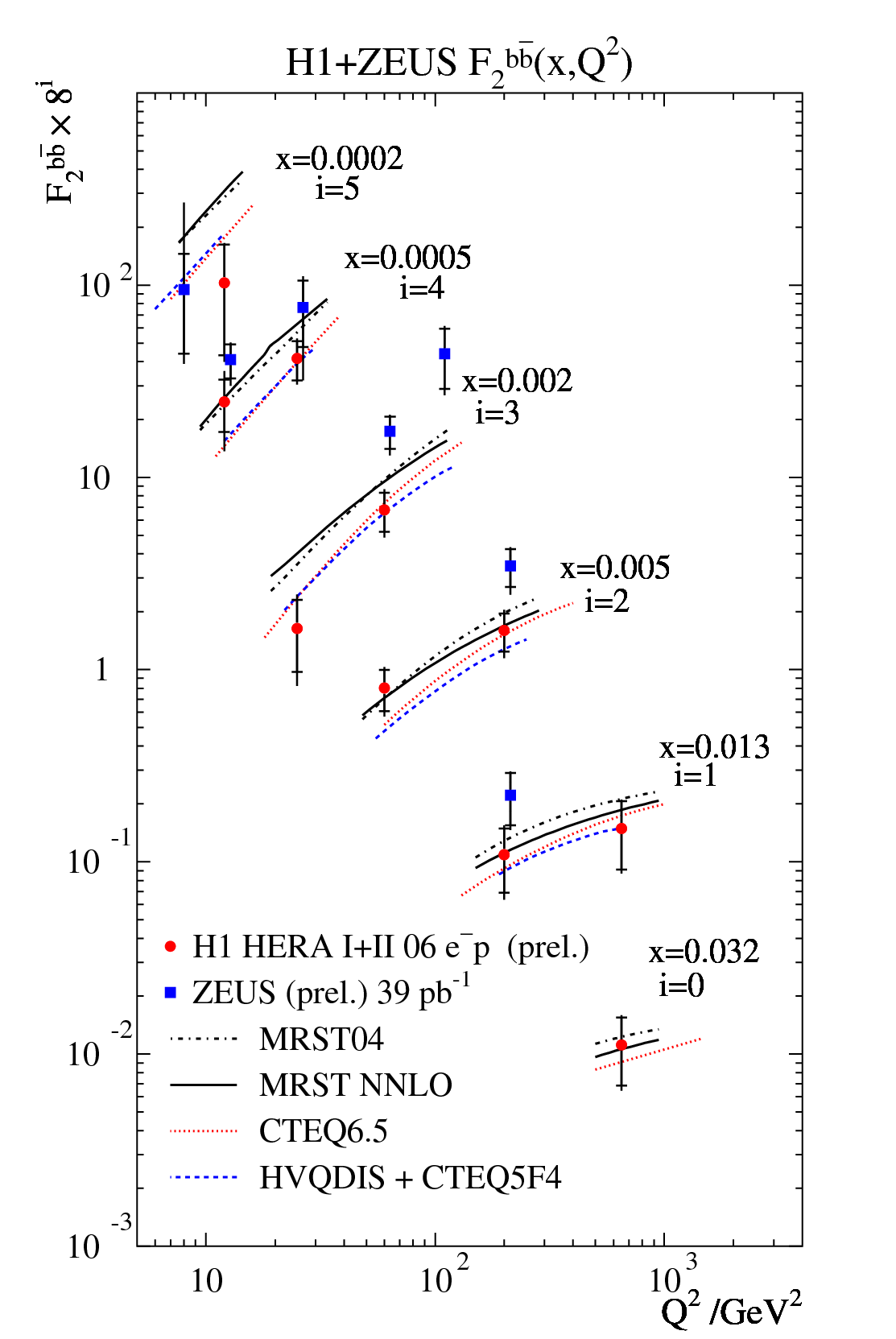}
  \caption{\Ftwob{} as a function of \Qsq{} for different $x$
    ranges. Also shown is a preliminary measurement from the ZEUS
    collaboration using the data from 2004. The inner errors bars
    show the statistical uncertainty, the outer error bars represent
    the statistical and systematic uncertainties added in
    quadrature. 
  }
  \label{fig:f2bc-f2b}
\end{figure}

From the reduced cross-section \Ftwoc{} can be extracted:  
\begin{equation*}
  \tilde{\sigma}^{\ccbar}(x, \Qsq) = \Ftwoc - 
  \frac{y^{2}}{(1 + (1-y)^{2})} \FLc\, .
\end{equation*}
The correction due to the longitudinal structure function, \FLc, is
small.
The same formulae with $c$ replaced by $b$
can be used to extract $\tilde{\sigma}^{\bbbar}$ and \Ftwob. 
The measurements of \Ftwoc{} and \Ftwob{} are shown in
Figures~\ref{fig:f2bc-f2c} and \ref{fig:f2bc-f2b}, respectively.
The rapid increase in \Ftwoc{} as a function of \Qsq{} at low $x$ is
clearly visible. The \Ftwob{} measurements are consistent with the same
trend, but are less precise.

\section{Conclusions}
\label{sec:conc}

\begin{figure}[htbp]
  \centering
  \includegraphics[width=0.85\columnwidth]{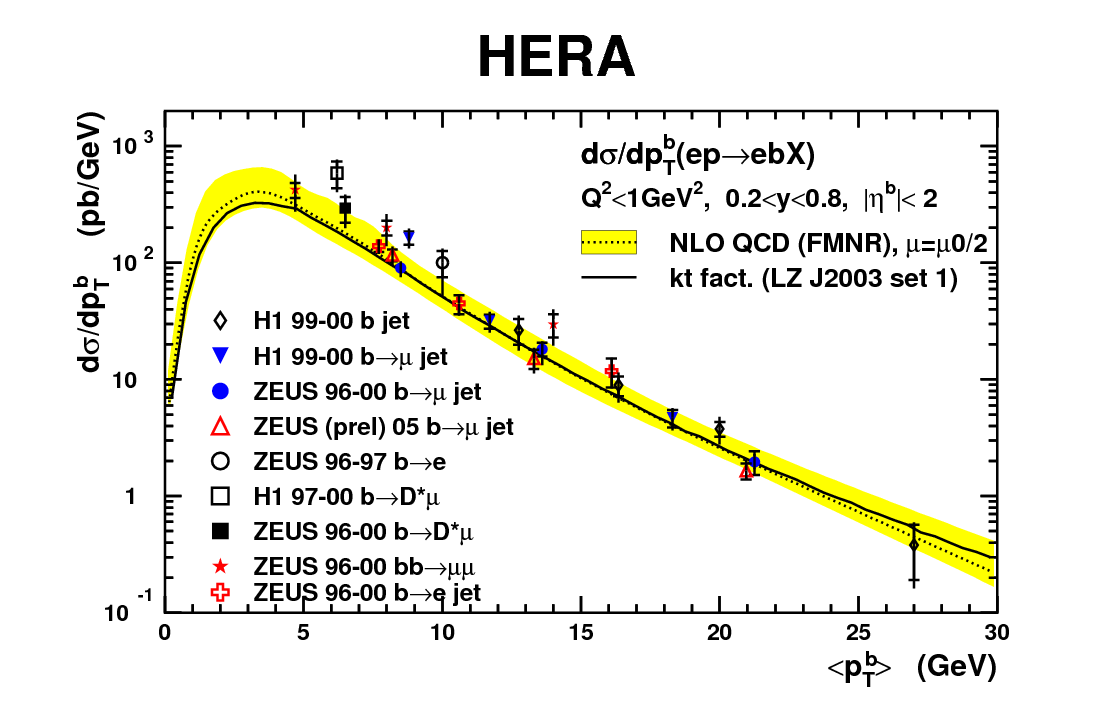}
  \caption{Differential cross section for $b$-quark production as a
    function of transverse momentum, \pTb, compared to the results of
    previous ZEUS measurements as indicated in the figure.  The
    measurements are shown as points. The inner error bar shows the
    statistical uncertainty and the outer error bar shows the
    statistical and systematic uncertainties added in quadrature.  The
    solid line shows the NLO QCD prediction from the FMNR program with
    the theoretical uncertainty shown as the shaded band.  }
  \label{fig:ptb}
\end{figure}

The many HERA measurements of beauty production in photoproduction are
compared in Figure~\ref{fig:ptb}.
The measurements presented here agree
well with the previous values, giving a consistent picture of
$b$-quark production in $ep$ collisions in the photoproduction regime,
and are well reproduced by the NLO QCD calculations.
For all the measurements leading order Monte Carlo predictions also
describe well the shapes of the distributions. 

There is a tendency for the charm measurements to overshoot the predictions
in the forward direction. Comparing charm quark predictions using
different PDFs shows that the cross-section is, as expected, sensitive
to the gluon PDF.

In the near future a number of final HERA~I measurements will be
published. With the HERA~II data the kinematic range of the
measurements can be extended and a combination of different tagging
methods should increase the precision of the measurements. The
improved HERA~II forward tracking will allow much improved studies of
heavy quark production in the forward direction.

\section*{Acknowledgements}
\label{sec:thanks}

It is a pleasure to thank the organisers for making this an
informative and enjoyable conference. I would like to thank Cristi
Diaconu, Achim Geiser, Markus Jüngst, Monica Turcato, André Schöning
and Matthew Wing for their help in preparing this talk.

{
\bibliographystyle{./elsarticle-num}
\raggedright\small
\bibliography{./brock_beach2008.bib}
}
\end{document}